\documentclass[runningheads]{llncs}
\usepackage[T1]{fontenc}
\usepackage{dsfont}
\usepackage{gensymb}
\usepackage{amssymb}
\usepackage{graphicx}
\begin{document}
\title{Diffusion Model-based FOD Restoration from High Distortion in dMRI\thanks{\scriptsize This work is supported by the National Institute of Health (NIH) under grants R01EB022744, RF1AG077578, RF1AG056573, RF1AG064584, R21AG064776, U19AG078109, and P41EB015922.}}
%
\titlerunning{FOD-Diffusion for Signal Loss Recovery}
%
\author{Shuo Huang\inst{1,2}, Lujia Zhong\inst{1,3} \and
Yonggang Shi\inst{1,2,3} \thanks{\scriptsize S Huang and L Zhong contributed equally to this work.}}
\authorrunning{S. Huang et al.}

\institute{Stevens Neuroimaging and Informatics Institute, Keck School of Medicine, University of Southern California (USC), Los Angeles, CA 90033, USA
\and
Alfred E. Mann Department of Biomedical Engineering, Viterbi School of Engineering, University of Southern California (USC), Los Angeles, CA 90089, USA
\and
Ming Hsieh Department of Electrical and Computer Engineering, Viterbi School of Engineering, University of Southern California (USC), Los Angeles, CA 90089, USA
\email{yonggang.shi@loni.usc.edu}}
\maketitle              
\begin{abstract}

Fiber orientation distributions (FODs) is a popular model to represent the diffusion MRI (dMRI) data. 
However, imaging artifacts  such as susceptibility-induced distortion in dMRI can cause signal loss and lead to the corrupted reconstruction of FODs, which prohibits successful fiber tracking and  connectivity analysis in affected brain regions such as the brain stem. Generative models, such as the diffusion models, have been successfully applied in various image restoration tasks. However, their application on FOD images poses unique challenges since FODs are 4-dimensional data represented by spherical harmonics (SPHARM) with the 4-th dimension exhibiting order-related dependency. In this paper, we propose a novel diffusion model for FOD restoration that can recover the signal loss caused by distortion artifacts. We use volume-order encoding to enhance the ability of the diffusion model to generate individual FOD volumes at all SPHARM orders. Moreover, we add cross-attention features extracted across all SPHARM orders in generating every individual FOD volume to capture the order-related dependency across FOD volumes. We also condition the diffusion model with low-distortion FODs surrounding high-distortion areas to maintain the geometric coherence of the generated FODs. We trained and tested our model using data from the UK Biobank ($n = 1315$). On a test set with ground truth ($n=43$), we demonstrate the high accuracy of the generated FODs in terms of root mean square errors of FOD volumes and angular errors of FOD peaks.  We also apply our method to a test set with large distortion in the brain stem area ($n=1172$) and demonstrate the efficacy of our method in restoring the FOD integrity and, hence, greatly improving tractography performance in affected brain regions.

\keywords{Fiber orientation distribution \and FOD restoration \and Susceptibility distortion \and Diffusion model.}
\end{abstract}

\section{Introduction}

\begin{figure}[tb]
\includegraphics[width = \textwidth]{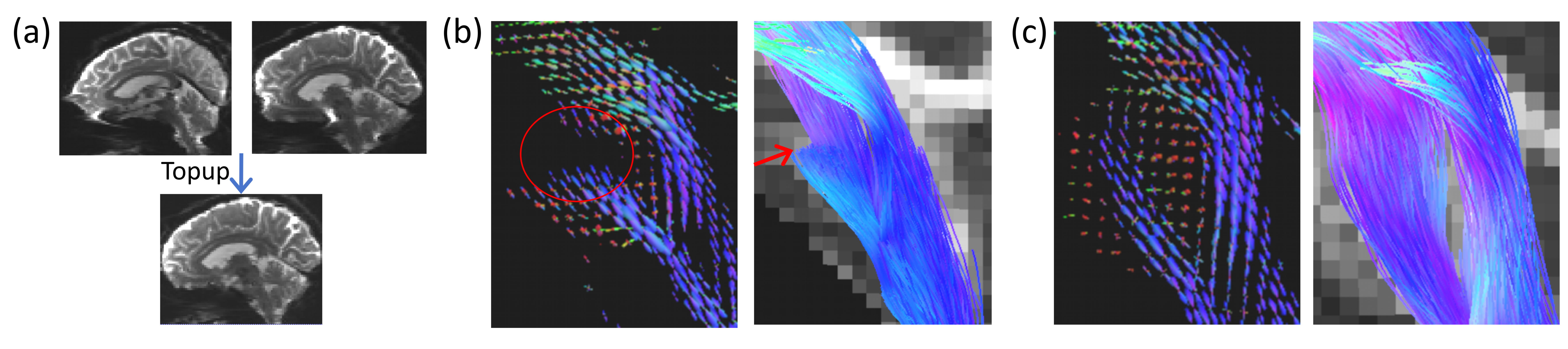} 
\centering
\caption{The residual distortion affects FODs and tractography. (a) The \textit{Topup} method from FSL \cite{topup} aligns the $b=0$ images from 2 opposite phase encoding directions to correct the susceptibility-induced distortion. (b) Corrupted FODs and failed tractography of the data in (a) illustrate the impact of the signal loss that cannot be recovered by distortion correction. (c) FODs from a low signal loss case and the successful fiber tracking results.} \label{Figure1}
\end{figure}

Fiber orientation distribution (FOD) is a popular representation \cite{TOURNIER20041176,fodcalculation} to model the configuration of fascicle trajectories based on high-resolution diffusion MRI (dMRI) and has been widely used in modern tractography techniques for the reconstruction of major fiber bundles \cite{Aydogan21}. Severe distortion artifacts, however, can prohibit the computation of valid FODs due to dMRI signal loss. While various distortion correction methods \cite{disco,topup,drcnet} were proposed to alleviate this problem, severe residual distortions are still widely present and pose a significant challenge to the successful reconstruction of fiber trajectories in regions with strong artifacts \cite{TANG2018227}(Fig .\ref{Figure1}). Building upon recent successes of diffusion models \cite{ddpm} in various image restoration and synthesis tasks \cite{diffusionmedical}, we propose in this work a novel diffusion model-based method for the generative recovery of FODs in brain regions with severe signal loss. 

\begin{figure}[tb]
\includegraphics[width = \textwidth]{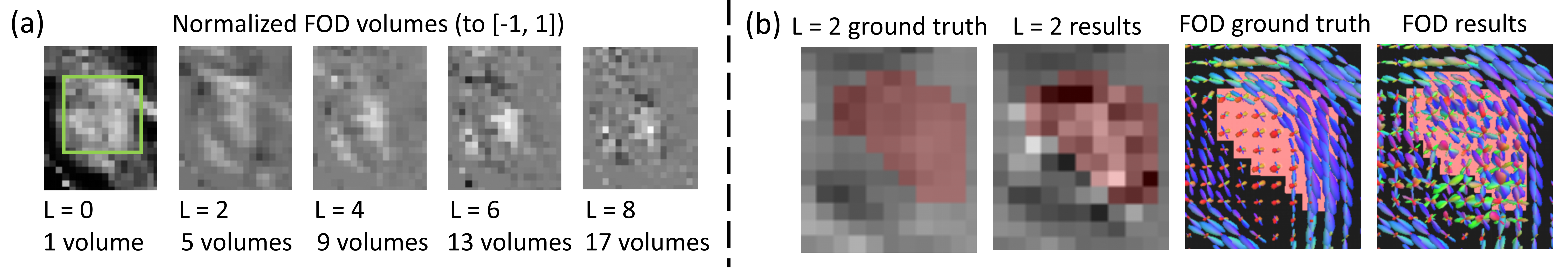}
\centering
\caption{The unconditioned DDPM model that only generates one FOD volume at each time failed in restoring the FODs. (a) The FODs have different numbers of volumes for each order. (b) Errors in individual volumes lead to erroneous FOD representations. } \label{Figure2}
\end{figure}



For the generative restoration \cite{diffusionrestoration,largesizerestoration,diffusionimpainting,diffusioninpaintinganother} and synthesis \cite{mrisynthesis,mrisynthesis3d} of natural images, denoising diffusion probabilistic models (DDPM) \cite{ddpm,diffusionmedical} have achieved great success because they were shown to be more stable in training and have more controllable generation procedures than other generative models such as GANs \cite{diffusionbeatgan} and VAEs \cite{diffusionwithvae}. The application of diffusion models for FOD restoration, however, poses unique challenges. First, FODs are typically represented by spherical harmonics (SPHARM) up to a maximum order $L$ \cite{fodcalculation}. The FODs are thus 4-dimensional images because they consist of multiple 3-D volumes at each SPHARM order. This would require large GPU memory and an enormous number of parameters if we generate all FOD volumes together. Second, while it is memory-efficient to treat all FOD volumes equally in training a generative DDPM model, the order-dependency of the FOD volumes can make it hard to ensure the validity of the generated FODs as shown in Fig \ref{Figure2}(a) and (b). Third, the generated FODs need to maintain geometric coherence with  surrounded voxels with low distortion and more reliable representation. 



In this work, we will develop a novel diffusion model-based method for the recovery of FODs in brain regions with high distortions, namely the \textit{FOD-Diffusion} model. We develop a volume order-aware diffusion model to make the FOD-Diffusion model suitable for generating  FOD volumes in both high- and low-frequency orders. We also provide a frequency order-balance cross-attention for extracting related information from all FOD volumes and helping the generation of individual FOD volumes.
Experiments on the large-scale UK Biobank dataset ($n = 1315$) \cite{UKBiobankcohol,ukbiobankdata} will demonstrate that our method successfully restores FODs from dMRI signal loss due to large residual distortion artifacts and hence greatly improved fiber tracking  through affected brain stem regions. 

\section{Method}
\begin{figure}[tb]
\includegraphics[width=\textwidth]{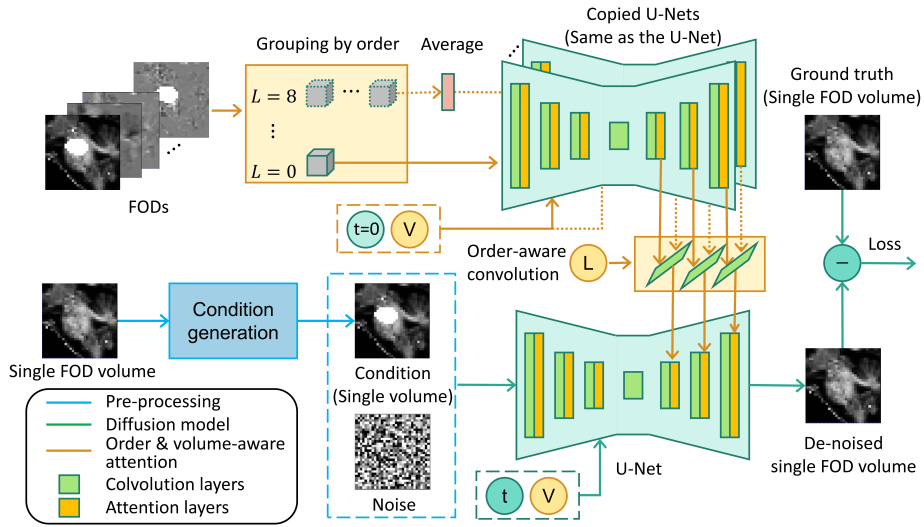}
\centering
\caption{The framework of the FOD-Diffusion model. Our FOD-Diffusion model takes the low-signal loss FODs as a condition in generation and uses the volume and order number encoding (denoted as V) to generate FODs in different frequency order. We also use low-signal loss FODs from all FOD volumes to extract the cross-attention information and help the generation of each FOD volume.} \label{Figure4}
\end{figure}

A detailed illustration of the proposed FOD-Diffusion model is shown in Fig. \ref{Figure4}. We will explain each aspect of the model in detail in the following sections.

\subsection{Order-aware Diffusion Model}%

We propose a volume order-aware diffusion model to enable the diffusion model to generate FODs across all different frequency orders.
Firstly, inspired by the time encoding in the diffusion model \cite{ddpm,attentiontime}, we propose a volume and frequency order-aware encoding (referred to as volume encoding) to enhance the model's ability to distinguish FODs of different frequencies and orientations. The volume encoding encodes an array $[L, V]$, where $L$ and $V$ are the frequency order number and the volume number of the single FOD volume, respectively.  

\begin{figure}[tb]
\includegraphics[width=3 in]{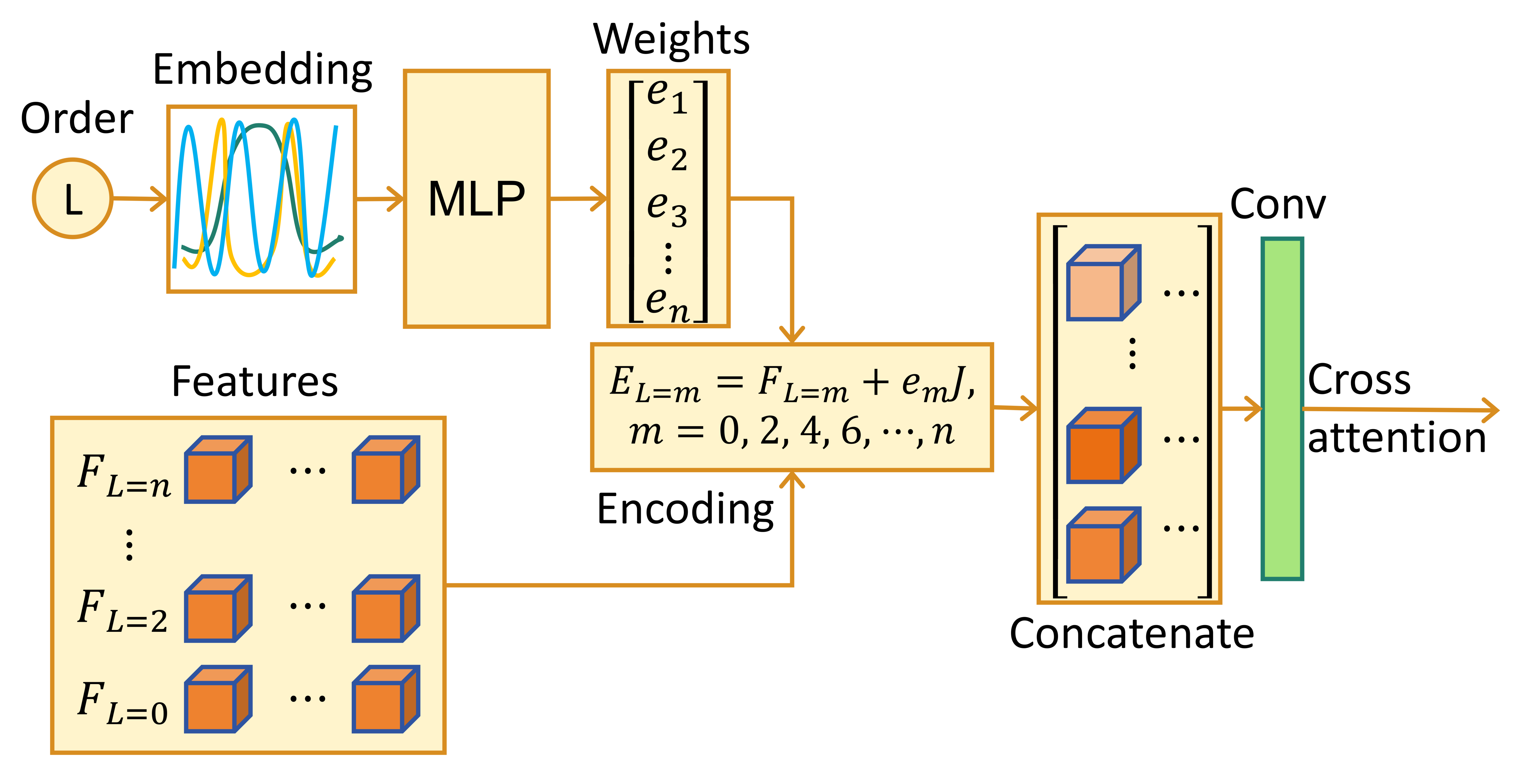}
\centering
\caption{Details of the frequency order-aware cross-attention calculation. We use the frequency order encoding to adjust the features from each order, and then we use a convolution layer to combine and select the features.} \label{Figure5}
\end{figure}

Moreover, we also use the low-signal loss regions 
as a condition to provide the model with sufficient constraints during generation. 
In the low-signal loss data, we assign high-signal loss regions to number 1 to distinguish them from the background.


We use the $L_1$ loss during training, and enhance the loss of the high-distortion regions,  
as shown in Eq. \ref{Equation1}:

\begin{equation}
    \mathcal{L} = 0.01 \times |\hat{x} - x|_1 + 0.99 \times \mathds{1}(|\hat{x} - x|_1),
\label{Equation1}
\end{equation}

\noindent where $\hat{x}$ and $x$ represent the generated FODs and the ground truth of the FODs, respectively. The indicator function $\mathds{1}(x) = x$ is applied within the mask of the high-signal-loss regions, while $\mathds{1}(x) = 0$ is applied in other voxels.

Since FODs at different orders have different gray-level ranges, they are normalized before generation. Components with $L = 0$ are normalized to the range $[0, 1]$, while other FODs are normalized to $[-1, 1]$. This ensures that the regions outside the brain remain 0 for all normalized FODs.

\subsection{Frequency-balanced Cross-attention}

The FODs represented by SPHARM have interdependence between different volumes, and we model the relationship between different FOD volumes to help the generation. We propose a frequency-balanced cross-attention method that extracts features from all FOD volumes and feeds them to the U-Net in the diffusion model through the cross-attention \cite{crossattention}. This method achieves balanced attention across each frequency order. 

As shown in Fig. \ref{Figure4}, for FOD volumes in each frequency order, we calculate their average to reduce the volume number to one in each frequency order. Specifically, for order $L = 0$, as it only has one volume, it can be directly used. Then, we use copied U-Nets from the diffusion model with time encoding $t = 0$ and volume encoding $V$ to extract features from the averaged FOD volumes in each frequency order. These features correspond to the output of the self-attention of each decoding block. Afterward, a frequency-order aware convolution with kernel size 1 is performed to select and combine the information that is most relevant to the FOD volume that needs to be generated.

We use copied U-Nets from the diffusion model's U-Net to extract cross-attention features. The parameters in copied U-Nets are frozen during the training. We copy the parameters to the copied U-Net when we update the U-Net in the diffusion model. This approach reduces the number of parameters that need to be trained; hence, it reduces the training time and prevents overfitting. 

More detail of the frequency order-aware convolution is shown in Fig. \ref{Figure5}. Firstly, we use sine and cosine functions to embed the order number. Afterward, a multilayer perceptron (MLP) layer is added to calculate the encoding weights $[e_1, e_2, ..., e_n]^T$. Eq. \ref{Equation2} is used to encode the features $F$ from each frequency order by adding the encoding weights:

\begin{equation}
    E_{L = m} = F_{L = m} + e_mJ,
\label{Equation2}
\end{equation}

\noindent where $m = 0, 2, ..., 8$ is the frequency order, and $J$ denotes the all-ones matrix.


The combined features from the 1×1 Conv are fed into the cross-attention through the Key and the Value in the U-Net of the diffusion model, and the Query is the feature from the single volume. 
The most useful information from other FOD volumes is selected from the Value to enhance the restoration of the current FOD volume under consideration \cite{crossattention}.



\section{Results and Discussion}
\subsection{Dataset and Preprocessing}
Our FOD-Diffusion method was trained, validated, and tested on the UK Biobank dataset. Overall, this study utilized $n=1315$ data points. They were extracted from the dataset randomly, with ages ranging from 30 years old to 80 years old. 

All extracted data were pre-processed using \textit{Topup} \cite{topup} and Eddy \cite{eddy,eddyhead} for correcting distortion, eddy current, and head motion. Then we calculated the FODs with the highest order $L=8$ using the method in Ref. \cite{fodcalculation}. To extract low-residual distortion cases for model training, we used the method in Ref. \cite{severitymap} to calculate the severity map of residual distortion, and we extracted 143 data that have both low mean signal loss ($< 0.25$) and high mean FOD integrity ($> 0.09$), which contains about 10\% of the data, for model training, validating and testing. We used $n = 90$ data for training, $n = 10$ data for validation, and $n = 43$ data for testing. We masked out regions in the brainstem of these data to act as the high-signal loss regions. The remaining $n=1172$ data that have high residual distortion formed a large test set to test the performance of the model in real data with high distortion. The masks of high-distortion regions were extracted based on the residual distortion severity map. 

\subsection{Model Training and Validation}




The model was trained for 100,000 iterations with the batch size of 8. We used AdamW optimizer in the training, with a learning rate of $10^{-5}$ initially and $10^{-6}$ after 70,000 iterations. We validated the model in every epoch, and the model with the least validation loss was selected. The training took about 68.5 hours on an NVIDIA RTX A5000 GPU, with about 22 GB of GPU memory. 
We used the DDPM for 1000 time steps in the inference, and the \textit{v-prediction} \cite{vprediction} was used in all experiments. The inference time for a single FOD volume was about 90 seconds.

\subsection{Ablation Studies}

\begin{figure}[tb]
\includegraphics[width=4 in]{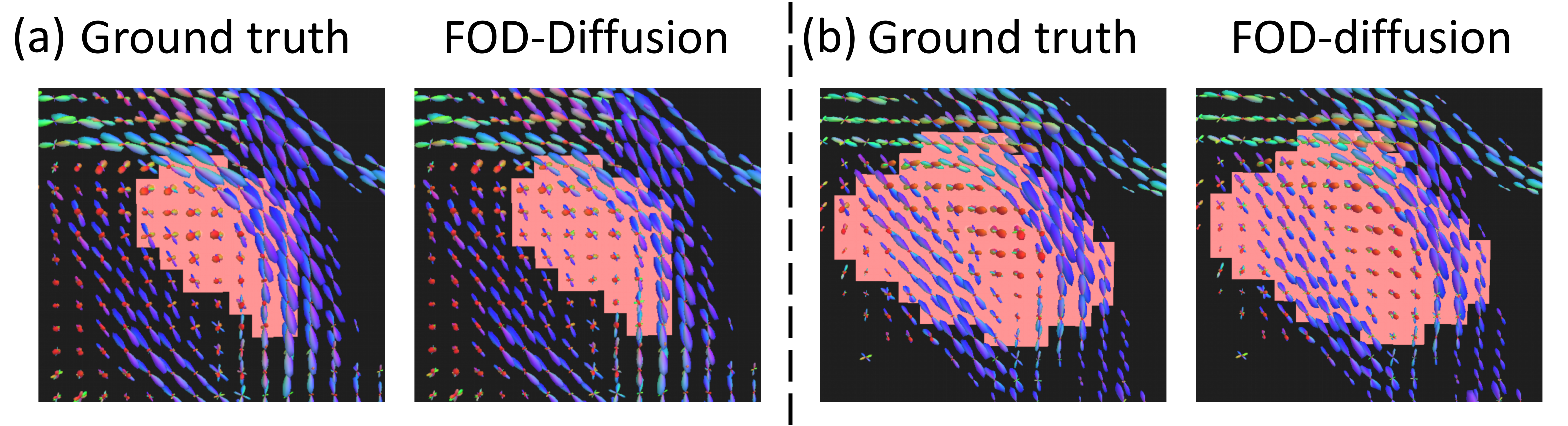} 
\centering
\caption{Examples of the ground truth FODs and the FODs of our FOD-Diffusion model from the test set. Especially, (a) is the same case in Fig. \ref{Figure2}} \label{Figure7}
\end{figure}

Fig. \ref{Figure7} shows two examples of the FODs for the ground truth and the results of our FOD-Diffusion model. These examples are from the test set. FODs generated by our FOD-diffusion model have high angular similarity with the ground truth. For the similarity of intensity, although some FODs have different intensities with the ground truth, the intensity is similar for most FODs. 

We then quantitatively compared our model with the unconditional DDPM and two ablation studies. The first ablation only takes the low-signal loss FOD volume (named vol) as the condition, and the second one also has the volume encoding (named enc). Table \ref{Table_root_mean_squared_error} shows the mean squared error of brain stem results from different methods. The unconditional diffusion model failed in FODs' generation.
Results of the ablation study show that all the improvements we added helped to improve the accuracy of the generated FODs.

We also compared the geometric differences between different methods, following the work in Ref. \cite{drcnet}. We first calculated the angles of the FOD peaks with the top three largest amplitudes that have peak values larger than 0.5 from both the ground truth FODs and the results FODs. Afterward, we calculated the corresponding smallest angular differences for the highest and second highest peak from the ground truth FODs (called ``$1^{st}$ peak'' and ``$2^{nd}$ peak'') to the peaks of the result FODs.
The results are shown in Table \ref{Table_angular_differences}. All ablation studies significantly overcome the unconditional diffusion model, and our FOD-Diffusion model has less angular differences and standard derivations than other models in the ablation studies.


\begin{table}[tb]
\centering
\caption{Root Mean Squared Errors of FODs in Test Set.}
\begin{tabular}{|l|l|l|l|l|l|l|}
\hline
Method & $L=0$ & $L=2$ & $L=4$ & $L=6$ & $L=8$ & FODs\\
\hline
Unconditional DDPM & 0.0976 & 0.0664 & 0.0453 & 0.0245 & 0.0114 & 0.0365\\
Diffusion + vol & 0.0136 & 0.0176 & 0.0143 & 0.0089 & 0.0046 & 0.0105\\
Diffusion + vol + enc & 0.0126 & 0.0152 & 0.0131 & \textbf{0.0088} & \textbf{0.0045} & 0.0097\\
FOD-Diffusion & \textbf{0.0121} & \textbf{0.0141} & \textbf{0.0129} & 0.0089 & \textbf{0.0045} & \textbf{0.0094}\\
\hline
\end{tabular}
\label{Table_root_mean_squared_error}
\end{table}



\begin{table}[tb]
\centering
\caption{Angular Differences for FODs in Test Set.}
\begin{tabular}{|l|l|l|}
\hline
Method & $1^{st}$ peak & $2^{nd}$ peak\\
\hline
Unconditional DDPM & $62.7 \degree \pm 2.1 \degree$ & $64.1 \degree \pm 2.1 \degree$ \\
Diffusion + vol & $2.2 \degree \pm 1.0 \degree$ & $7.5 \degree \pm 1.5 \degree$\\
Diffusion + vol + enc & $2.1 \degree \pm 0.9 \degree$ & $7.3 \degree \pm 1.4 \degree$\\
FOD-diffusion & $\textbf{2.0} \degree \pm \textbf{0.8} \degree$ & $\textbf{7.2} \degree \pm \textbf{1.4} \degree$\\
\hline
\end{tabular}
\label{Table_angular_differences}
\end{table}



\subsection{Performance on Data with High Distortion}
Because we do not have ground truth for high distortion data, we will use the group-wise distribution on the FODs' integrity at the pons regions to evaluate the FOD-Diffusion model in the high signal loss UK Biobank data ($n = 1172$).  Here we compare our FOD-Diffusion method with the \textit{Topup} method, as it is a representative registration-based method and the most widely used method in connectome research. 

\begin{figure}[tb]
\includegraphics[width = 4 in]{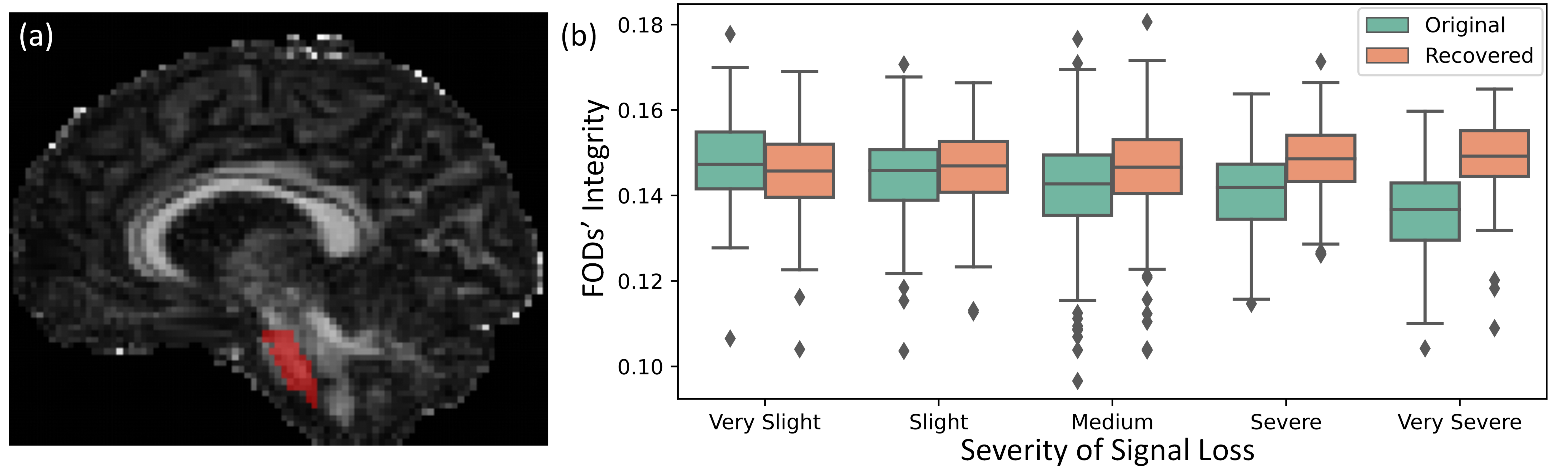}
\centering
\caption{Distributions of FODs' integrity at the pons region before and after signal loss recovery for high residual distortion data ($n = 1172$). (a) shows the position of the pons' masks, (b) shows the distribution of FODs' integrity.} \label{Figure6}
\end{figure}

Fig. \ref{Figure6} shows the distribution of the FODs' integrity at the pons region before and after the signal loss recovery. The 1172 data are evenly divided into five groups based on the severity of signal loss, named groups of ``very slight'', ``slight'', ``medium'', ``severe'', and ``very severe'' residual distortion, respectively.  
Each of the first four groups has 236 subjects, and the last group has 228 subjects. The mean intensity of the $L = 0$ component at the pons ROIs (the atlas is defined in Ref. \cite{jhuatlas}) in Fig. \ref{Figure6}(a) was calculated and used as a measure of FOD integrity in the evaluation following the work in Ref. \cite{severitymap}. This evaluation method is efficient because the intensity of the $L = 0$ component represents the mean FOD value \cite{fodcalculation}, and the mean FOD decreases with the increase of the residual distortion.
Fig. \ref{Figure6} (b) shows that while the FOD integrity of the original FODs reduces with the increase of the severity in signal loss, the integrity  of restored FODs have similar distributions across the groups. This shows that our method can recover signal loss for data with high residual distortions. 

\begin{figure}[tb]
\includegraphics[width=\textwidth]{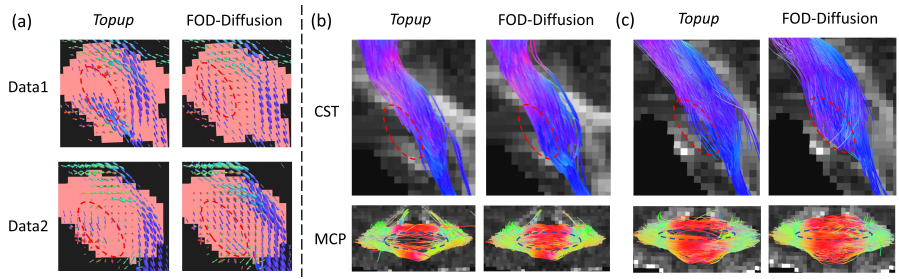}
\centering
\caption{Tractography of CST and MCP for 2 high signal loss cases. (a) shows the FODs of the \textit{Topup} and FOD-Diffusion methods. (b) and (c) are the tractography results of data1 and data2, respectively. ``Data1'' is the same data in Fig. \ref{Figure1}(b).} \label{Figure8}
\end{figure}

The trachography results for two high signal loss cases are shown in Fig. \ref{Figure8}. Here, we calculated the corticospinal tract (CST) \cite{tractcst} and the middle cerebellar peduncle (MCP) \cite{tractmcp} for both the \textit{Topup} results and the results of our FOD-Diffusion method. These two fiber bundles share the same region of interest at the pons (the red mask in Fig. \ref{Figure6}(a)) for generating the seeds. Fig. \ref{Figure8} shows that the tractography results of CST using FOD-Diffusion cover more pons region, and the MCP generated using FOD-Diffusion can cover the whole pons region. Therefore, our FOD-Diffusion method can successfully restore FOD-based fiber connectivity at the pons region.

\section{Conclusion}
In this work, we proposed a diffusion model-based method for FOD restoration to recover the signal loss caused by high residual distortions.
For the generation of complex 4D FOD data, our model is memory efficient that can be trained on one GPU with 24GB GPU memory. 
We demonstrated the performance of our method in brainstem regions using data from UK Biobank. In data with ground truth (low distortion), we quantitatively validated the accuracy of our generated FODs. For data with high distortion, we demonstrated the restoration of FOD integrity and the potential of restored FODs in helping fiber tracking of important brainstem pathways. 

Future work will include testing our method in more datasets from clinical studies, such as Alzheimer’s Disease Neuroimaging Initiative (ADNI) \cite{adni} and   Health \& Aging Brain among Latino Elders (HABLE) \cite{hable}. We will also test our method in more complex brain regions with high residual distortions such as the temporal lobe.

\subsubsection{Acknowledgements} Authors thank Miss Mahsa Torfeh and Miss Yi Liu from University of Southern California for the help in editing the grammar of this work. We also appreciate Mr. Wenhao Chi, Mr. Jianwei Zhang, Mr. Zhiwei Deng, Miss Jiaxin Yue and Dr. Xinyu Nie for the useful discussion and suggestions.



\end{document}